# Dynamic critical behavior of multi-grid Monte Carlo for two-dimensional nonlinear $\sigma$-models

Gustavo Mana[a], Tereza Mendes[a,*], Andrea Pelissetto[b] and Alan D. Sokal[a]

[a]Department of Physics, New York University, 4 Washington Place, New York, NY 10003, USA

[b]Dipartimento di Fisica and INFN, Università degli Studi di Pisa, Pisa 56100, Italy

We introduce a new and very convenient approach to multi-grid Monte Carlo (MGMC) algorithms for general nonlinear $\sigma$-models: it is based on embedding an $XY$ model into the given $\sigma$-model, and then updating the induced $XY$ model using a standard $XY$-model MGMC code. We study the dynamic critical behavior of this algorithm for the two-dimensional $O(N)$ $\sigma$-models with $N = 3, 4, 8$ and for the $SU(3)$ principal chiral model. We find that the dynamic critical exponent $z$ varies systematically between these different asymptotically free models: it is approximately 0.70 for $O(3)$, 0.60 for $O(4)$, 0.50 for $O(8)$, and 0.45 for $SU(3)$. It goes without saying that we have no theoretical explanation of this behavior.

Multi-Grid Monte Carlo (MGMC) [1–8] is a collective-mode approach that introduces block updates (of fixed shape but variable amplitude) on all length scales. The basic ingredients of the method are:

1) **Interpolation operator:** A rule specifying the shape of the block update. The interpolations most commonly used are *piecewise-constant* and *piecewise-linear*.

2) **Cycle control parameter $\gamma$:** An integer number that determines the way in which the different block sizes are visited. In general, blocks of size $2^l$ are updated $\gamma^l$ times per iteration. Thus, in the W-cycle ($\gamma = 2$) more emphasis is placed on large length scales than in the V-cycle ($\gamma = 1$).

3) **Basic (smoothing) iterations:** The local Monte Carlo update that is performed on each level. Typically one chooses to use *heat-bath* updating if the distribution can be sampled in some simple way, and *Metropolis* otherwise.

4) **Implementation:** The computations can be implemented either in the *recursive multi-grid* style using explicit coarse-grid fields [9,1–4], or in the *unigrid* style using block updates acting directly on the fine-grid fields [10,5–8]. For a $d$-dimensional system of linear size $L$, the computational labor per iteration is

$$\mathrm{Work(MG)} \sim L^d \qquad \text{if } \gamma < 2^d \tag{1}$$

for the recursive multi-grid approach[2], and

$$\mathrm{Work(UG)} \sim \begin{cases} L^d \log L & \text{for } \gamma = 1 \\ L^{d+\log_2 \gamma} & \text{for } \gamma > 1 \end{cases} \tag{2}$$

for the unigrid approach. Thus, the unigrid implementation is marginally more expensive for a V-cycle, but prohibitively more expensive for a W-cycle.

The efficiency of the method can be analyzed rigorously in the case of the Gaussian (free-field) model, for which it can be proven [2,11] that critical slowing-down is completely eliminated. That is, the *dynamic critical exponent* $z$ (the exponent with which the *autocorrelation time* $\tau$ diverges as the correlation length $\xi$ tends to infinity) is zero.[3] More precisely, the algorithm with piecewise-linear interpolation exhibits $z = 0$ for both V-cycle and W-cycle, while the one with piecewise-constant interpolation has $z = 0$ only for the W-cycle (the piecewise-constant V-cycle has $z = 1$).

*Speaker at the conference.

[2]For the W-cycle in $d = 1$ there appears an extra factor $\log L$.

[3]See [12] for a pedagogical discussion of the various autocorrelation times and their associated dynamic critical exponents.

One is therefore motivated to apply MGMC to "nearly Gaussian" systems: one might hope that critical slowing-down would be completely eliminated (possibly modulo a logarithm) or at least greatly reduced compared to the $z \approx 2$ of local algorithms. In particular, we are interested in applying MGMC to asymptotically free two-dimensional $\sigma$-models, such as the $N$-vector models [also called $O(N)$-invariant $\sigma$-models] for $N > 2$, the $SU(N)$ principal chiral models, and the $RP^{N-1}$ and $CP^{N-1}$ $\sigma$-models.

In view of the rigorous results for the Gaussian case, we want to investigate two questions:

1) Is $z = 0$ for all asymptotically free two-dimensional $\sigma$-models? If not, does $z$ vary from one asymptotically free model to another?

2) Is the algorithm with piecewise-constant interpolation and a W-cycle as efficient as the one with piecewise-linear interpolation and a V-cycle, i.e. is $z_{PC,W}$ equal to $z_{PL,V}$ for these models?

The key design choice in a MGMC algorithm is that of the interpolation operator; indeed, this choice determines most of the remaining ingredients. If one chooses a "smooth" interpolation such as piecewise-linear, then it is usually impossible to implement true recursive MGMC[4], as there is no simple form for the induced coarse-grid Hamiltonians. Therefore one is obliged to use the unigrid style, a V-cycle, and Metropolis updating. We call this the "German" approach [5–8]. On the other hand, if one chooses a "crude" interpolation such as piecewise-constant, then is obliged to use a W-cycle in order to have a chance at $z < 1$; but piecewise-constant interpolation usually gives rise to a simple coarse-grid Hamiltonian (typically a slight generalization of the fine-grid Hamiltonian), so that one can use the recursive multi-grid style and, at least in principle, heat-bath updating. This is the approach taken by our group [1–4].

The "German" version has the advantage of being easy to implement for diverse models, but its use of Metropolis updates introduces several free parameters that have to be adjusted, making it more difficult to test systematically. "Our" version has no free parameters, but its implementation is cumbersome and model-dependent, in the sense that the program (and in particular the heat-bath subroutine) has to be drastically rewritten for each distinct model. For example, among the $N$-vector models, the only ones that can be handled conveniently are $N = 2$ [3] and $N = 4$ [4], by exploiting the isomorphism with the $U(1)$ and $SU(2)$ groups, respectively.

With this problem in mind, we have developed a new implementation of MGMC that combines some of the advantages of both methods; in particular, it can be used conveniently for a large class of $\sigma$-models with very little modification of the program. The idea is to *embed* angular variables $\{\theta_x\}$ into the given $\sigma$-model, and then update the resulting induced $XY$ model by our standard (piecewise-constant, W-cycle, heat-bath, recursive) MGMC method. We do not claim that this approach is superior in practice to the "German" method — that remains to be determined — but we do think that it is well suited for the systematic study of the dynamic critical behavior of MGMC algorithms.

For the models discussed here, the induced $XY$ Hamiltonian is of the form

$$\mathcal{H}_{embed} = -\sum_{\langle xx' \rangle} [\alpha_{xx'} \cos(\theta_x - \theta_{x'}) + \beta_{xx'} \sin(\theta_x - \theta_{x'})] \ , \qquad (3)$$

where the induced couplings $\{\alpha_{xx'}, \beta_{xx'}\}$ are given in terms of the values of the original spins. We illustrate this for the two cases thus far studied:

1) $N$-vector models: The original variables are unit vectors in $\mathbb{R}^N$, and the original Hamiltonian is

$$\mathcal{H}_{N-vector} = -\beta \sum_{\langle xx' \rangle} \boldsymbol{\sigma}_x \cdot \boldsymbol{\sigma}_{x'} \ . \qquad (4)$$

The embedding is given by choosing randomly a plane $P$ in $\mathbb{R}^N$, and defining $\theta_x$ to be the angular coordinate of the projection of $\boldsymbol{\sigma}_x$ onto $P$. (At each iteration a new random plane is chosen.) If each vector $\boldsymbol{\sigma}_x$ is decomposed into its parts parallel and perpendicular to the plane, $\boldsymbol{\sigma}_x = \boldsymbol{\sigma}_x^{\parallel} + \boldsymbol{\sigma}_x^{\perp}$, then the induced $XY$ Hamiltonian is of the form

[4] In particular, this is the case for the nonlinear $\sigma$-models.

(3) with

$$\alpha_{xx'} \;=\; \beta\, |\sigma_x^{\parallel}|\, |\sigma_{x'}^{\parallel}| \tag{5}$$

$$\beta_{xx'} \;=\; 0 \tag{6}$$

Though the couplings $\{\alpha_{xx'}\}$ are random, the induced $XY$ model is ferromagnetic if the original $N$-vector model is.

2) $SU(N)$ principal chiral models: The original variables are group elements $U_x \in SU(N)$, and the Hamiltonian is given by

$$\mathcal{H}_{SU(N)} \;=\; -\beta \sum_{\langle xx' \rangle} \mathrm{Re}\,\mathrm{tr}(U_x^{\dagger}\, U_{x'})\; . \tag{7}$$

In this case we choose to write the embedding from a different point of view, namely we define the updated variable $U_x^{new}$ by

$$U_x^{new} \;=\; R\, e^{i\theta_x T}\, R^{-1}\, U_x^{old}\; , \tag{8}$$

where $R$ is an $SU(N)$ matrix picked at random at each step (the analogue of the random plane in the previous case), and $T \in \mathfrak{su}(N)$ is a fixed traceless diagonal matrix with entries $\pm 1$ or 0 [we use $T = \mathrm{diag}(1, -1, 0, \ldots, 0)$]. $T$ satisfies

$$e^{i\theta T} \;=\; T^2 \cos\theta \,+\, iT \sin\theta \,+\, (I \,-\, T^2)\; . \tag{9}$$

Therefore, the induced $XY$ Hamiltonian is of the form (3), with couplings

$$\alpha_{xx'} \;=\; \beta\, \mathrm{Re}\,\mathrm{tr}(U_x^{\dagger}\, R\, T^2\, R^{-1}\, U_{x'}) \tag{10}$$

$$\beta_{xx'} \;=\; \beta\, \mathrm{Im}\,\mathrm{tr}(U_x^{\dagger}\, R\, T\, R^{-1}\, U_{x'}) \tag{11}$$

The embedded model is simulated with initial condition $\theta_x = 0$ (i.e. $U_x^{new} = U_x^{old}$ for all $x$). Note that the couplings in this case are not only disordered but are in general frustrated.

We have investigated the dynamic critical behavior of the MGMC method (using piecewise-constant interpolation, W-cycle and heat-bath updates) for a variety of $\sigma$-models in two dimensions. A few years ago, a careful study [4] of the direct MGMC method for the 4-vector model showed $z_{int,\mathcal{M}^2} = 0.60 \pm 0.07$ (compared to $z \approx 2$ for conventional local algorithms such as heat bath and Metropolis). Therefore, critical slowing-down is greatly reduced but not completely eliminated. Indeed, it is not completely eliminated even for the *one*-dimensional version of the same model, for which our data [13] are consistent with a logarithmic growth of the autocorrelation time.

Recently, we have applied the $XY$-embedding algorithm to the $N$-vector model for $N = 3, 4, 8$, finding that:

1) For $N = 4$, the embedding algorithm belongs to the same dynamic universality class as the direct MGMC algorithm, as expected.

2) The dynamic critical exponent $z$ is $N$-dependent: we have $z_{int,\mathcal{M}^2} \approx 0.70, 0.60, 0.50$ for $N = 3, 4, 8$, respectively (error bars in each case are roughly $\pm 0.05$). It thus appears that $z$ decreases as $N$ gets larger; and it might conceivably be the case that $z$ tends to zero as $N$ tends to infinity, which would be consistent with the vague idea that the $N = \infty$ model is "essentially Gaussian".

The method was also tested for the $SU(3)$ principal chiral model, yielding the preliminary estimate $z = 0.45 \pm 0.02$.

These exponents are found by fitting to the *dynamic finite-size-scaling Ansatz*

$$\tau_{int,A}(\beta, L) \;\approx\; \xi(\beta, L)^{z_{int,A}}\, g_A(\xi(\beta, L)/L)\; , \tag{12}$$

where $A$ is an observable (here the square of the magnetization, which is found to be the slowest mode among those we study), $\tau_{int,A}(\beta, L)$ is its *integrated autocorrelation time* [12], $\xi(\beta, L)$ is the *second-moment correlation length* described in [4], and $g_A$ is a smooth function. We plot $\tau_{int,\mathcal{M}^2}(\beta, L)/\xi(\beta, L)^z$ as a function of $\xi(\beta, L)/L$, and vary $z$ until the points fall nicely on a single curve (except for possible corrections to scaling for the smaller lattices). Figures 1, 2 and 3 show the results for the $O(3)$, $O(8)$ and $SU(3)$ models, using our preferred choices for the exponents $z$. The corrections to scaling appear to be weaker for $SU(3)$ than for $O(3)$ or $O(8)$; we don't know why.

This work was supported in part by NSF grant DMS-9200719.

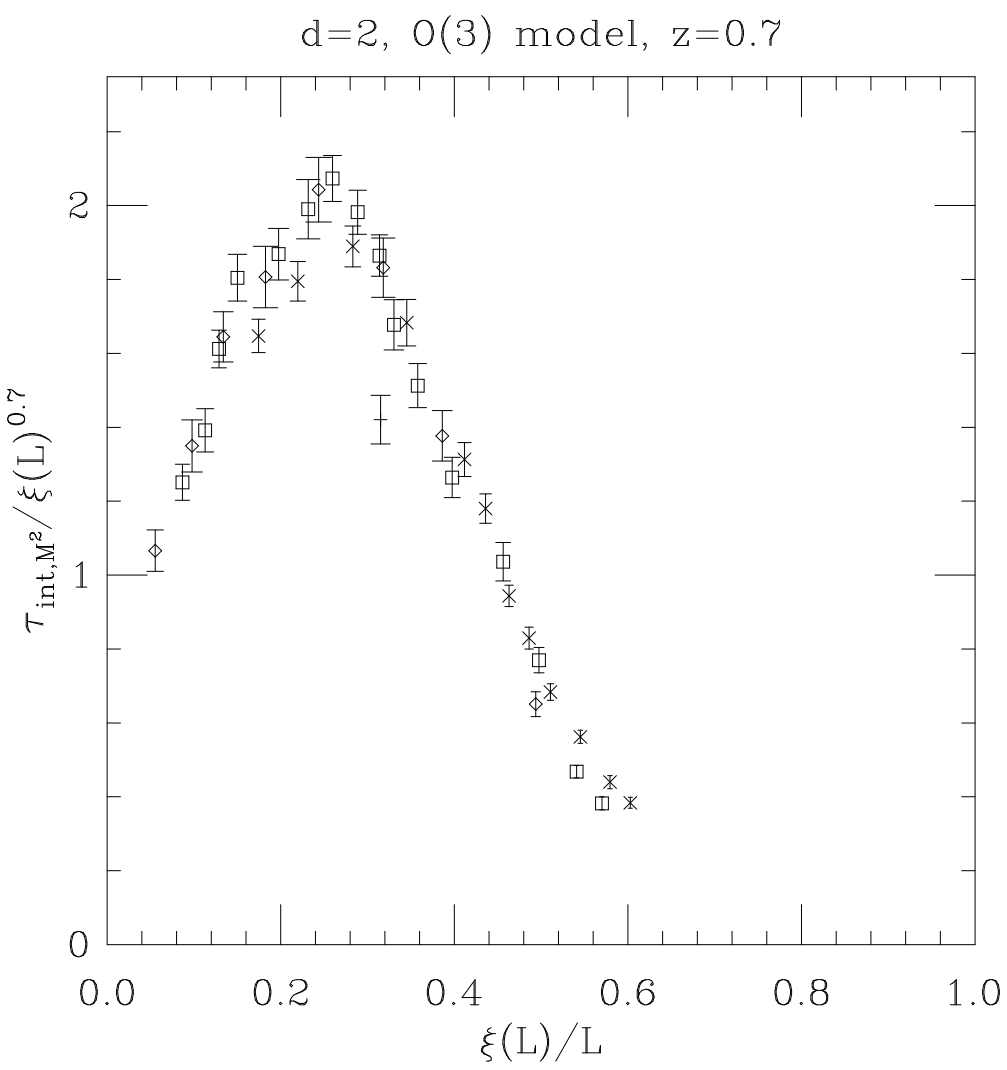

Figure 1. $\tau_{int,\mathcal{M}^2}(\beta, L)/\xi(\beta, L)^z$ versus $\xi(\beta, L)/L$ for the $O(3)$ model, with $z = 0.70$. Lattice sizes $L$ are 32 (+), 64 (×), 128 (□), 256 (◇).

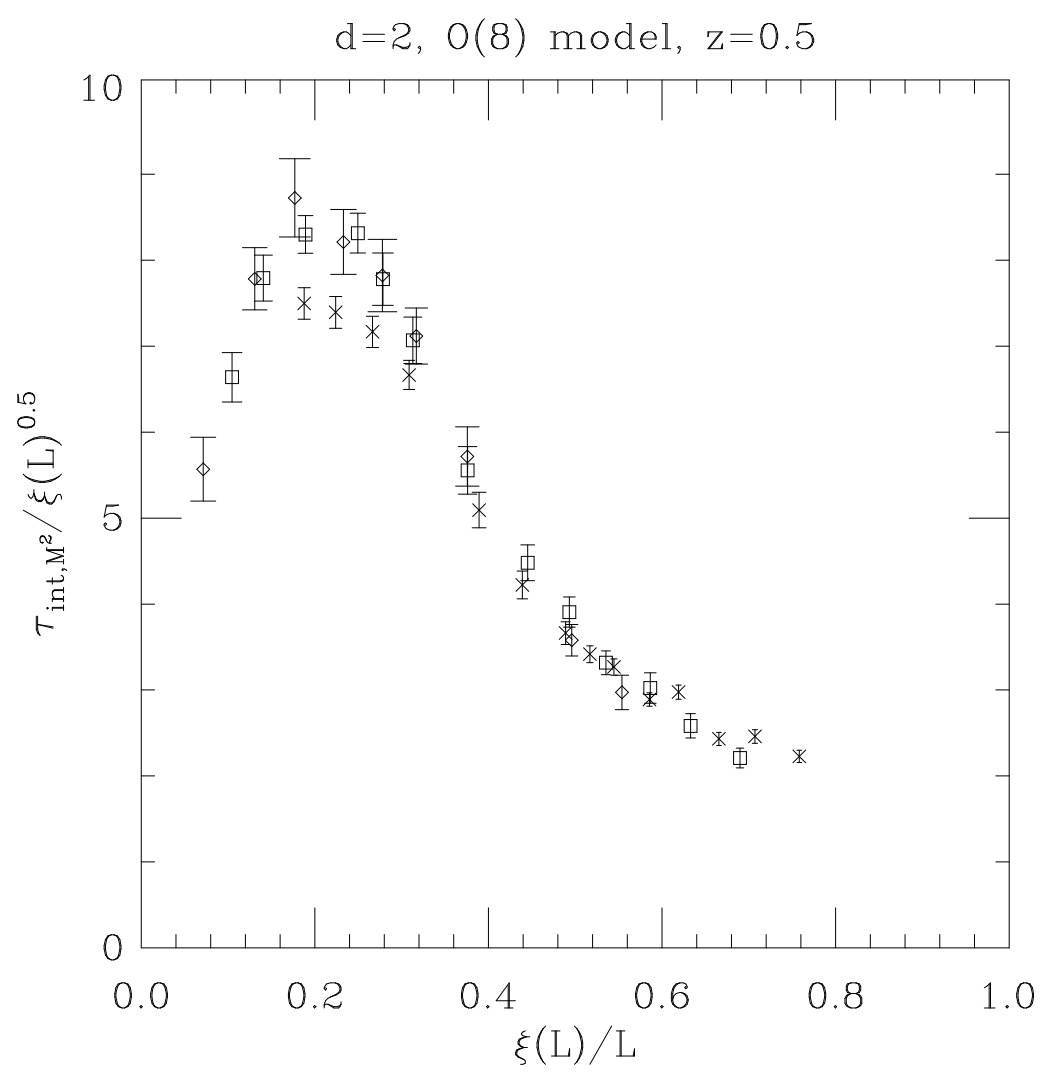

Figure 2. Same, for the $O(8)$ model, with $z = 0.50$.

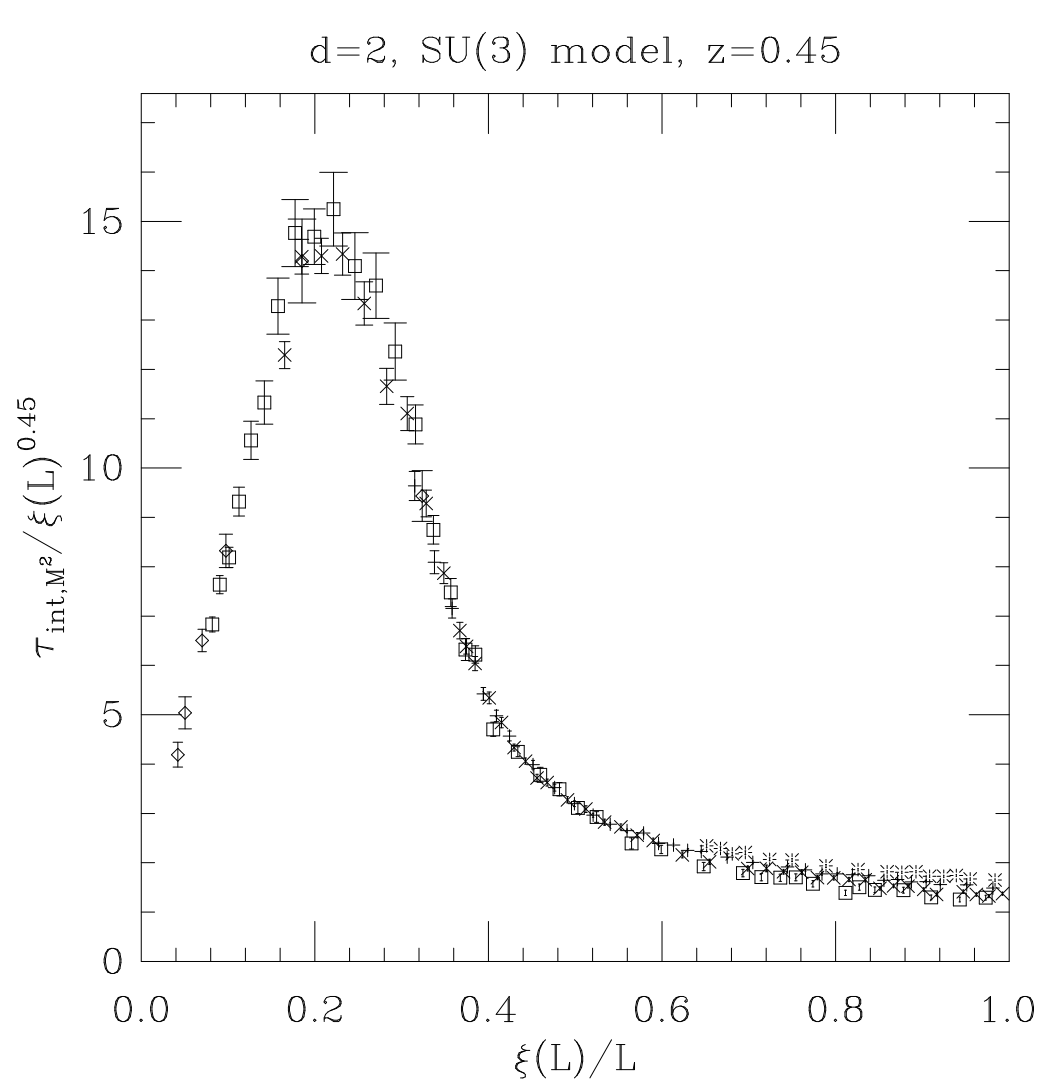

Figure 3. Same, for the $SU(3)$ model, with $z = 0.45$. Includes also $L = 16$ (∗).